\documentclass[a4paper]{article}

\usepackage{amsmath}
\usepackage{amssymb}
\usepackage{bm}
\DeclareMathOperator*{\argmin}{argmin}

\usepackage[pdftex]{graphicx}
\graphicspath{{./figs/}}
\DeclareGraphicsExtensions{.pdf,.jpeg,.png}

\usepackage{booktabs} 

\usepackage{INTERSPEECH_v2}

\title{Speaker identification from the sound of the human breath}
\name{Wenbo Zhao, Yang Gao, Rita Singh}
\address{Carnegie Mellon University, Pittsburgh, USA}
\email{\{wzhao1,yanggao,rsingh\}@cs.cmu.edu}

\begin{document}

\maketitle
\begin{abstract}
This paper examines the speaker identification potential of breath sounds in continuous speech. Speech is largely produced during exhalation. In order to replenish air in the lungs, speakers must periodically inhale. When inhalation occurs in the midst of continuous speech, it is generally through the mouth. Intra-speech breathing behavior has been the subject of much study, including the patterns, cadence, and variations in energy levels. However, an often ignored characteristic is the {\em sound} produced during the inhalation phase of this cycle. Intra-speech inhalation is rapid and energetic, performed with open mouth and glottis, effectively exposing the entire vocal tract to enable maximum intake of air. This results in vocal tract resonances evoked by turbulence that are characteristic of the speaker's speech-producing apparatus. Consequently, the sounds of inhalation are expected to carry information about the speaker's identity. Moreover, unlike other spoken sounds which are subject to active control, inhalation sounds are generally more natural and less affected by voluntary influences. The goal of this paper is to demonstrate that breath sounds are indeed bio-signatures that can be used to identify speakers. We show that these sounds by themselves can yield remarkably accurate speaker recognition with appropriate feature representations and classification frameworks. 
%
%
%
\end{abstract}

\section{Introduction}
Intervocalic breath sounds are fundamentally different from relaxed breath sounds that occur outside of speech. This is because breath plays an important role in controlling the dynamics of speech. Natural speech is produced as a person exhales. \textit{It is almost impossible to produce sustained speech during inhalation}  \cite{warren1976aerodynamics}. As a person speaks, a specific volume of air is pushed out through the lungs and trachea into the vocal chambers, gated through the vocal folds in the glottis.
Intervocalic breath sounds happen when the speaker exhausts the volume of air previously inhaled during continuous speech, and needs to inhale again.
This inhalation is generally sharp and rapid, and volumetrically anticipatory of the next burst of speech. The volume of air inhaled also depends on the air-intake capacity of the speaker's nasal and oral passageways, trachea and inner structures leading to the lungs, and further varies with myriads of other factors that relate to the speaker's lung capacity, energy levels, muscular agility, etc. Since exhalation is volumetrically linked to inhalation, by association the quality of the speech produced during exhalation also varies with all of these factors. Furthermore, when a person inhales, the vocal tract is usually lax, and is in its natural shape. The lips are not protruded, nor do the articulators obstruct the vocal tract in any way. In lax configurations, differences between speakers are expected to show up prominently as differences in resonant frequencies (due to differences in facial skeletal proportions and dimensions of the vocal chambers), differences in relative sound intensities (due to different lung capacities and different states of health), etc.

For all these reasons, we expect that many parameters of the speaker's persona have their effects, and possibly measurable signatures, embedded in intervocalic breath sounds. Our goal in this paper is to experimentally show that these person-specific signatures within human breath (inhalation) sounds can indeed be used for speaker identification.

\begin{figure}[!ht]
\centering
\begin{minipage}[b]{0.8\linewidth}
\includegraphics[width=0.92\linewidth]{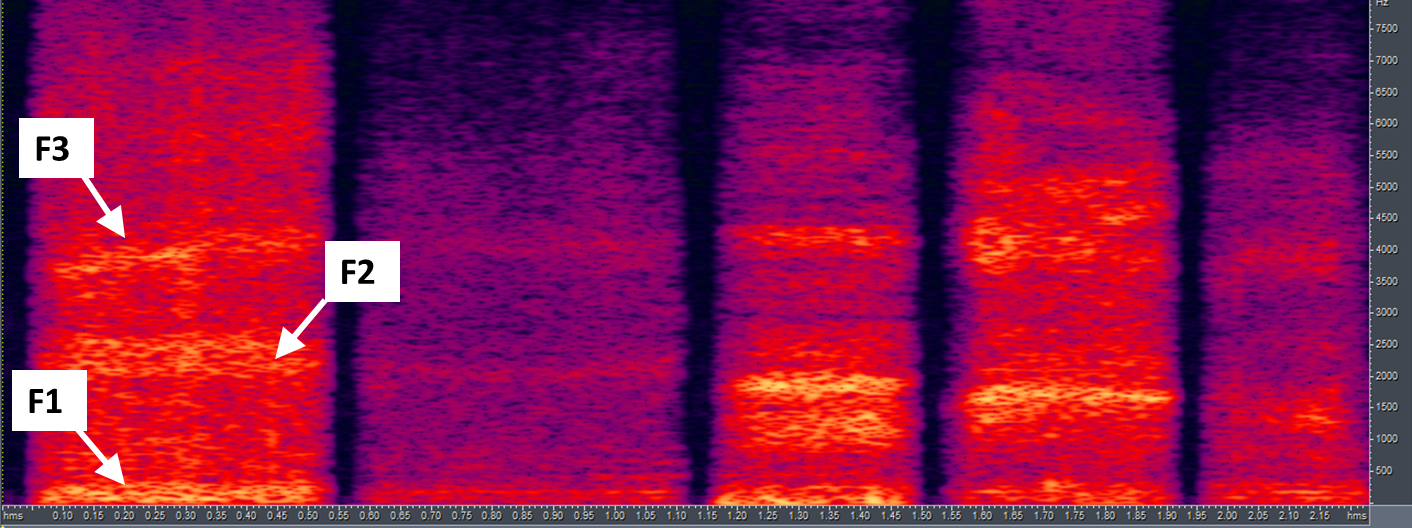} 
(a)
\end{minipage}\hfill
\centering
\begin{minipage}[b]{0.8\linewidth}
\includegraphics[width=0.92\linewidth]{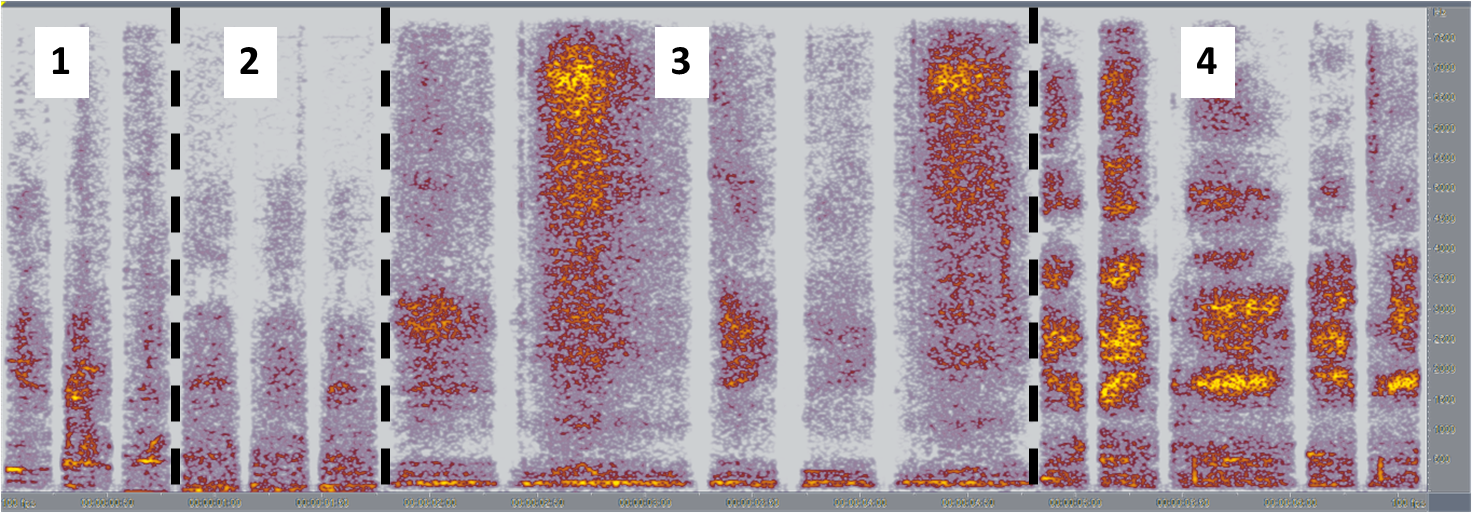} 
(b)
\end{minipage}\hfill
\caption {\textbf{(a)} Spectrogram of breath sounds of a 4-year old child during continuous speech. The formants F1, F2 and F3 correspond to the resonance of breath sounds, and are clearly visible. \textbf{(b)} Breath sounds of Mr. Donald Trump (label 3), and of his impersonators (Labels 1, 2 and 4). All signals were energy-normalized and are displayed on exactly the same scale. }
\label{fig:breathsounds}
\end{figure}

This can be useful in many real-life scenarios, especially those of forensic importance. For example, we have shown in an earlier study   \cite{singhmeskimen2016}, that breath is invariant under disguise and impersonation. Its resonance patterns are normally not under voluntary control of the speaker, and in general are extremely difficult to modify in a consistent manner for mechanical or cognitive reasons. Fig. \ref{fig:breathsounds} shows the spectrograms of the breath sounds of a child and four adult speakers, three of who were attempting to impersonate the fourth speaker. This example is in fact extracted from the public performances of voice artists who were attempting to impersonate the US presidential candidate in the 2016 elections in USA -- Mr. Donald Trump. We see the qualitative differences in the breath sounds and durations clearly in these examples. Even though in reality the impersonators of Mr. Trump sound very similar, their breath sounds show very distinctive speaker-specific patterns.

\vspace{-0.5cm}

\subsection{Prior work} Speaker identification from speech signals is a widely applied and well researched area, with decades of work supporting it. Technology from this area has been the mainstay of forensic analysis of voice as well, an area that has been largely centered around the topics of speaker identification  \cite{meuwly2009forensic, champod2000inference, rose2003forensic, campbell2009forensic}, verification  \cite{becker2008forensic, reynolds2000speaker}, detection of media tampering, enhancement of spoken content, and profiling  \cite{singhprofiling, singhmipro2016}. All of these areas have used articulometric considerations to advantage. However, there are no reported studies that use or suggest the use of breath sounds strongly in \textit{any} of these forensic contexts. The closest application -- and one that takes a stretch of imagination to relate to forensics -- in fact comes from the medical field, where the sound of the patient's breath is sometimes used very subjectively by the clinician to deduce the patient's medical condition (such as lung function, respiratory diseases, response to their treatment, etc.)  \cite{loudon1988volumes, henderson1965temporal, winkworth1995breathing}.

The rest of this paper is organized as follows. In Section \ref{sec:feats} we present a very brief review of two feature formulations that we derive from breath sounds for our experiments. In section \ref{sec:classifiers}, we describe a problem formulation and solution framework for speaker identification based on convolutional neural networks (CNN) with Long-Short Term Memory (LSTM). In section \ref{sec:expts}, we present our experiments, and in Section \ref{sec:concl} we present our conclusions.

\section{Brief review of feature formulations}\label{sec:feats}
Breath is a turbulent sound.
Since the spectral distributions of turbulent sounds are well represented by Gaussians, we hypothesize that the standard speaker identification techniques based on supervectors and i-vectors \cite{dehak2011front} would work with breath sounds as well. 

In this section we describe two feature formulations that we selected for representing breath sounds. The two formulations we chose are i-vectors, the currently accepted best features used in commercially deployed state-of-art speaker identification and verification systems  \cite{dehak2011front}, and a set of novel CNN-RNN based features derived from constant-Q representations of the speech signal. We describe these briefly in the subsections below. 

\subsection{I-Vectors as features for breath sounds}
Identity-vector (i-vector) based feature representations are ubiquitously used in state-of-art speaker identification and verifications systems.

In order to obtain i-vectors for any speech recording, the distribution of Mel-frequency Cepstral Coefficient (MFCC) vectors derived from it is modeled as a Gaussian mixture. The parameters of this Gaussian mixture models (GMM) are in turn obtained through maximum {\em a posteriori} adaptation of a {\em Universal Background Model} that represents the distribution of all speech  \cite{reynolds1997comparison,reynolds2015universal}. The mean vectors of the Gaussian in the adapted GMM are concatenated into an extended vector, known as a GMM  \textit{Supervector}, which represents the distribution of the MFCC vectors in the recording  \cite{kinnunen2010overview}.

I-vectors are obtained through factor analysis of GMM supervectors. Following the factor analysis model, each GMM supervector $\mathbf{M}$ is modeled as $\mathbf{M} = \mathbf{m} + \mathbf{T}\mathbf{w_M}$, where $\mathbf{m}$ is a global mean, $\mathbf{T}$ is a triangular loading matrix comprising bases representing a {\em total variability space}, and $\mathbf{w_M}$ is the i-vector corresponding to $\mathbf{M}$. The loading matrix $\mathbf{T}$ and mean $\mathbf{m}$ are learned from training data through the Expectation Maximization (EM) algorithm  \cite{moon1996expectation}. Subsequently, given any recording $\mathbf{M}$, its i-vector can also be derived using EM.

\subsection{Constant-Q spectrographic features}
For our purposes, a constant-Q spectrogram of a speech signal may be derived by computing the constant-Q spectra of consecutive frames of the signal, which are then displayed in temporal succession on the spectrogram. As in general spectrographic representation used for speech signals, the frames span durations of $20 \sim 30$ ms and overlap by 50\%-75\%.

For each frame $s[t]$, we compute its constant-Q transform $\mathbf{x}_t^{\mathrm{cq}}$  \cite{brown1991calculation}. In the case of digital recordings, specifically, the constant-Q transform of each frame $s[t]$ of the signal is given by
\begin{align}
x^{\mathrm{cq}}[k] = \tfrac1{N_k} \sum_{n < N_k} s[n]w_{N_k}[n]e^{-j2\pi nQ/N_k},\,\, k = 1,\ldots,K
\label{eq:const_q}
\end{align}
with the sampling frequency $f_s$, the minimum frequency $f_0$, the maximum frequency $f_{\mathrm{max}}$, and the number of filters per octave $b$. Then the number of frequency bins $K=b \log_2 \frac{f_\mathrm{max}}{f_0}$, the $k^{\mathrm{th}}$ center frequency $f_k = f_0 2^{\frac{k}{b}}$, and the bandwidth of the $k^{\mathrm{th}}$ filter $\delta f_k^{\mathrm{cq}} = f_k(2^{\tfrac1{b}}-1)$. Therefore, the ratio of $f_k$ to $\delta f_k^{\mathrm{cq}}$ is $Q = (2^{\tfrac1{b}}-1)^{-1}$, and the window length for the $k^{\mathrm{th}}$ bin $N_k = Q\frac{f_s}{f_k}$.
Collecting all the $K$ coefficients gives $\mathbf{x}_t^{\mathrm{cq}} = \left[x^{\mathrm{cq}}[1], \ldots, x^{\mathrm{cq}}[K]\right]$. Concatenating $\mathbf{x}_t^{\mathrm{cq}}$ for $T$ frames into a matrix $\mathbf{X}$ gives us the constant-Q spectrogram for input signal. We use these spectrograms as features in our CNN-LSTM experiments.

The primary reason for the choice of constant-Q spectrograms is that the variations in the spacings of the harmonics due to variations in pitch on a normal spectrogram become constant shifts in frequency on a constant-Q spectrogram. In addition, the filters used in constant-Q computation have geometrically spaced center frequencies and bandwidths like MFCC, allowing better discrimination for speech sounds in general.

\section{CNN-LSTM based approach}\label{sec:classifiers}
In this section, we first describe how we formulate the speaker identification problem, and then describe the CNN-LSTM based framework as the solution to the speaker identification problem.

\subsection{Problem formulation}
Consider a collection of $N$ constant-Q spectrograms $\mathbb{X} = \{\mathbf{X}_1, \ldots, \mathbf{X}_N\}$, $\mathbf{X}_i \in \mathbb{R}^{F \times T}$, $i=1, \ldots, N$, where $F$ is the number of frequency bins, and $T$ is the number of frames. Denote the collection of the corresponding labels as $\mathbb{Y} = \{\mathbf{y}_1, \ldots, \mathbf{y}_N\}$, $\mathbf{y}_i \in \mathbb{R}^{C}$, $i=1,\ldots, N$, where $C$ is the number of speaker classes. The label $\mathbf{y}_i$ for speaker class $c$ has the $c^{\mathrm{th}}$ entry being $1$ and the rest entries being $0$s. Given the speech spectrograms $\mathbf{X}$, our goal is to design a classifier $h: \mathbb{R}^{F \times T} \to \mathbb{R}^{C}$ to predict the probability mass over the $C$ classes: $\widehat{\mathbf{y}} = h(\mathbf{X}) = \mathbb{P}(\mathbf{y} \mid \mathbf{X})$. We first define the loss function

\begin{align}
L(\widehat{\mathbf{y}}, \mathbf{y}) &= D_{\mathrm{KL}}(\mathbf{y} \parallel \widehat{\mathbf{y}}) = \sum_{j=1}^{C} y_j \log{\frac{y_j}{\widehat{y}_j}}
\label{eq:loss}
\end{align}
where $D_{\mathrm{KL}}(\mathbf{y} \parallel \widehat{\mathbf{y}})$ is the Kullback–-Leibler distance between the true probability mass $\mathbf{y}$ and the predicted probability mass $\widehat{\mathbf{y}}$. It is often used to measure the distance between densities. Now we define the classification risk
\begin{align}
R(h) &= \mathbb{E}_{(\mathbb{X}, \mathbb{Y}) \sim \mathcal{D}} \left[ L(\widehat{\mathbf{y}}, \mathbf{y}) \right] 
= \mathbb{E}_{(\mathbb{X}, \mathbb{Y}) \sim \mathcal{D}}\left[D_{\mathrm{KL}}(\mathbf{y} \parallel \widehat{\mathbf{y}})\right]
\label{eq:risk}
\end{align}
which is the expectation of the loss (\ref{eq:loss}) over the data distribution $\mathcal{D}$. Since the true data distribution is unknown, we consider the empirical risk instead
\begin{align}
\widehat{R}(h) &= \tfrac1{N} \sum_{i=1}^{N} D_{\mathrm{KL}}(\mathbf{y}_i \parallel \widehat{\mathbf{y}}_i) 
= \tfrac1{N} \sum_{i=1}^{N} \sum_{j=1}^{C} y_{ij} \log{\frac{y_{ij}}{\widehat{y}_{ij}}}
\label{eq:emp_risk}
\end{align}
which is the average of the losses over all the $N$ samples. Such follows our optimization objective of minimizing the empirical risk (\ref{eq:emp_risk})
\begin{align}
h^* = \argmin_{h \in \mathcal{H}}\widehat{R}(h)
\label{eq:obj}
\end{align}
where $\mathcal{H}$ is a class of classification functions. Later in this section we construct $\mathcal{H}$ as multi-layer neural networks.

\subsection{Speaker identification framework}
\begin{figure}[!ht]
\centering
\begin{minipage}[b]{0.8\linewidth}
\includegraphics[width=0.92\linewidth]{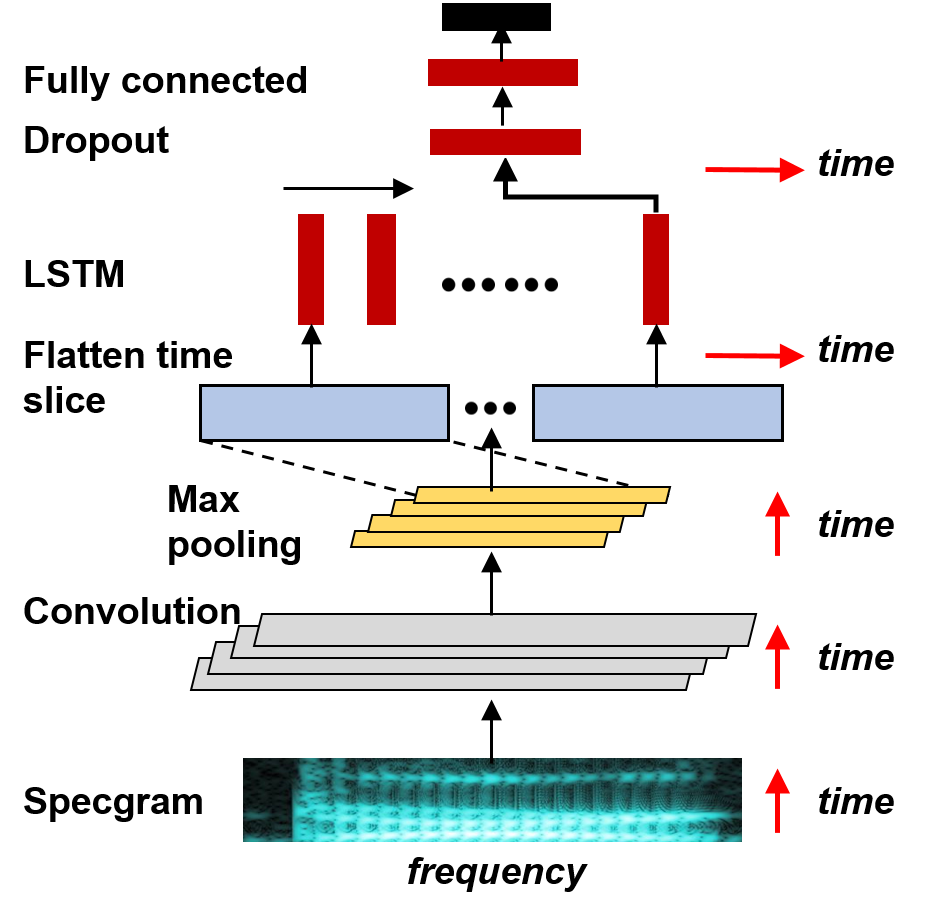} 
\end{minipage}\hfill
\centering
\begin{minipage}[b]{0.5\linewidth}
\includegraphics[width=0.92\linewidth]{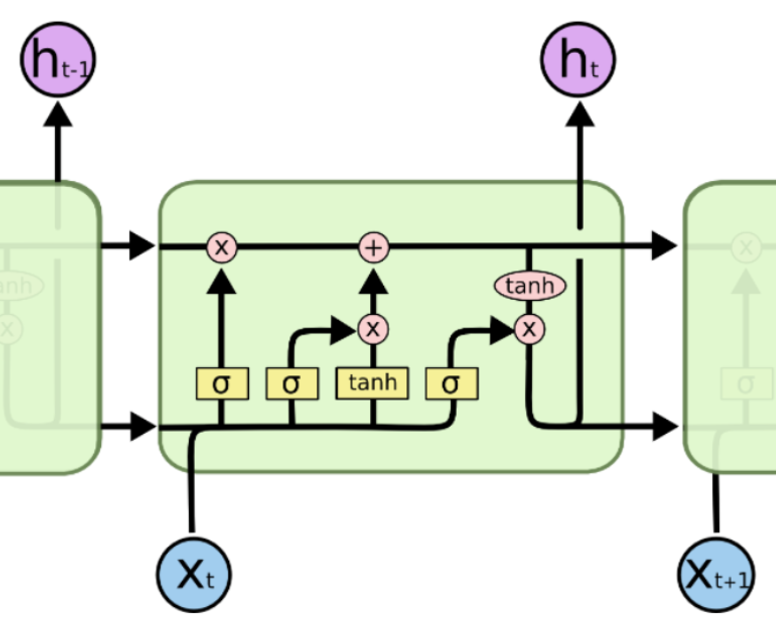} 
\end{minipage}\hfill
\caption{\textbf{Top:} CNN-LSTM network architecture. \textbf{Bottom:} Structure of an LSTM unit. $x_t$ and $h_t$ are the input and output respectively at the $t^{\mathrm{th}}$ step, and $\sigma$ is the logistic function.}
\label{fig:net}
\end{figure}

Next, we describe the speaker identification network. Fig.~\ref{fig:net} shows an overview of the network. It is composed of a convolutional layer, a max-pooling layer, a LSTM layer, a dropout layer, and a fully connected layer.

The convolutional layer convolutes a same set of filters with an input spectrogram to learn its shift-variant features, which can be used to identify speakers. Specifically, for a spectrogram $\mathbf{X} \in \mathbb{R}^{F \times T}$, the convolutional layer convolutes it with $L$ filters $\mathbf{W}_i \in \mathbb{R}^{U \times V}$, $i=1, \ldots, L$ with size $(U, V)$. The resulting feature map $\mathbf{Z}_i^{\mathrm{conv}} = \mathbf{X} \ast \mathbf{W}_i + b_i$, where $b_{i}$ is a bias term. The feature map is then activated $\mathbf{X}_i^{\mathrm{conv}} = r(\mathbf{Z}_i^{\mathrm{conv}})$ with the rectifier function $r(x) = \max(0, x)$.

The collection of feature maps $\mathbb{X}^{\mathrm{conv}} = \{\mathbf{X}_1^{\mathrm{conv}}, \ldots, \mathbf{X}_L^{\mathrm{conv}}\}$ are then down-sampled along frequency with max-pooling mechanism \cite{zeiler2014visualizing} to reduce the amount of parameters and computation in the network, and hence to also control over-fitting. In order to learn the temporal information between the frames in each feature map, the down-sampled feature maps $\mathbb{X}^{\mathrm{pool}}$ are flattened along time and fed into a LSTM layer. The LSTM layer consists of a sequence of memory units to selectively remember the past sequence information. For a single unit, as illustrated in the bottom panel of Fig.~\ref{fig:net}\cite{Olah2015lstm}, it has one memory cell and three control gates: the forget gate, the input gate, and the output gate \cite{hochreiter1997long}. We take the output from the last time step.

The output of the LSTM layer $\mathbf{h}$ is further fed into a dropout layer to combat over-fitting \cite{srivastava2014dropout}. The resultant output $\mathbf{h}^{\mathrm{drop}}$ is finally passed to a fully connected layer and normalized using soft-max function
\begin{align}
\widehat{y}_i = \frac{e^{\mathbf{w}_i^T \mathbf{h}^{\mathrm{drop}}}}{\sum_{j=1}^{C}e^{\mathbf{w}_j^T \mathbf{h}^{\mathrm{drop}}}},\,\, i = 1,\ldots,C
\label{eq:likelihood}
\end{align}
where $\mathbf{w}_i$ is the weight in fully connected layer. This final output $\widehat{\mathbf{y}} = \left [\widehat{y}_1, \ldots, \widehat{y}_C \right]$ is the multi-class likelihoods for the $C$ speakers. We can then minimize the risk (\ref{eq:obj}), where the function class $\mathcal{H} = \{h(\mathbf{w})\}$ for all the parametrization of $\mathbf{w}$ in the network. Substitution (\ref{eq:likelihood}) into (\ref{eq:obj}) yields
\begin{align}
\mathbf{w}^* &= \argmin_{\mathbf{w}} \widehat{R}(h(\mathbf{w})) \nonumber \\
&= \argmin_{\mathbf{w}} \tfrac1{N}\sum_{i=1}^{N} (\log \sum_{j=1}^{C}e^{\mathbf{w}_j^T\mathbf{h}^{\mathrm{drop}}} - \mathbf{w}_i^T \mathbf{h}^\mathrm{drop})
\label{eq:net_obj}
\end{align}
To obtain the optimal set of parameters $\mathbf{w}^*$, we train our network using back-propagation. The gradients of the risks with regard to the parameters in the last layer are back-propagated and the parameters are updated until convergence.

\section{Experimental results}
\label{sec:expts}
We first describe our experimental setup and results with i-vectors, and then those with the CNN-LSTM speaker identification network that we propose in Section \ref{sec:classifiers}. The data used for both sets of experiments were the same. We used Sphinx-3, a state-of-art Hidden Markov Model (HMM) based automatic speech recognition (ASR) system to first obtain accurate phoneme segmentations for all the speech included in the LDC Hub-4 1997 Broadcast news database  \cite{hub4-97}. The database comprises single-channel recordings of read speech from multiple news anchors and people interviewed within the news episodes. The recordings are sampled at 16000 Hz. The ASR system was trained on this  database, and the acoustic models obtained were used to obtain highly accurate phoneme segmentations. Breath was modeled as a phoneme during the training process, and thus the process of phoneme segmentation directly yielded the breath sounds that we needed for our experiments. 

The complete set of breath sounds extracted from the Hub-4 database included more than 3000 combinations of speaker, channel (broadband and telephone), fidelity (high, low, medium) and type of speech (read and conversational), of which we only chose breath sounds that corresponded to high fidelity clean read speech signals for our experiments. Since the goal of this paper is confined to demonstrating that breath can indeed be used to identify speakers, we did not attempt to explore of performance in different speech styles, channel types, noise conditions, etc. In the following subsection, we describe the specific experiments we conducted in detail.

\subsection{I-vector based experiments}
The breath data used in i-vector experiments comprised 9915 instances corresponding to 50 speakers. For each speaker, 70\% of the data were used as training set and 30\% as test set. 

For generating i-vectors, we first used mixtures of 512 Gaussians to generate supervectors. The universal background model for generating the supervectors was trained using all the training data. The computed i-vectors were normalized by centering the data and normalizing the length  \cite{i-vectorN}. For both the test data and training data, i-vectors of different dimensions (20, 30, 40, 60, 80, 100, 200, 300) were computed, and speaker recognition accuracies were obtained using each of these feature sets. In one set of experiments, performed with SVMs, we reduced the dimensionality of all vectors with initial dimensionality 60 and above, down to 49 using Linear Discriminant Analysis (LDA)  \cite{izenman2013linear}.

Our experiments used two types of classifiers: a multi-class SVM, and a two-layer neural network. The multi-class SVM was implemented using the lib-svm library  \cite{LIBSVM}. The neural network is implemented using Theano  \cite{2016arXiv160502688short} and Keras  \cite{chollet2015keras}. The architecture of the 
multi-layer neural network included different activation functions in different layers. None were used in the input layer, the first hidden layer used the rectified linear unit (ReLu) and the second hidden layer, when present, used the Sigmoid activation function. The soft-max function was used in the output layer. The training learning rate was 0.1 with a decay rate of $1e^{-9}$, and the momentum was 0.9. The batch size used in each epoch of training was 400. 

\begin{figure}[!ht]
\begin{minipage}[b]{0.9\linewidth}
\centering
\includegraphics[width=0.9\linewidth]{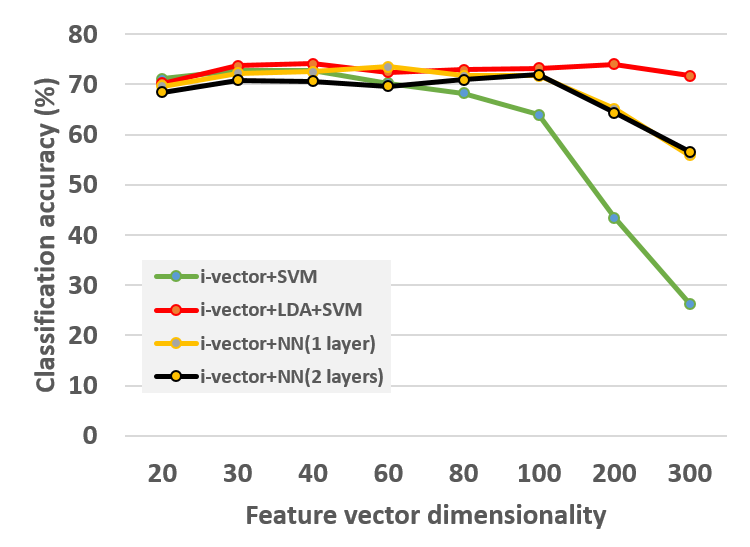} \hfill
\includegraphics[width=0.8\linewidth]{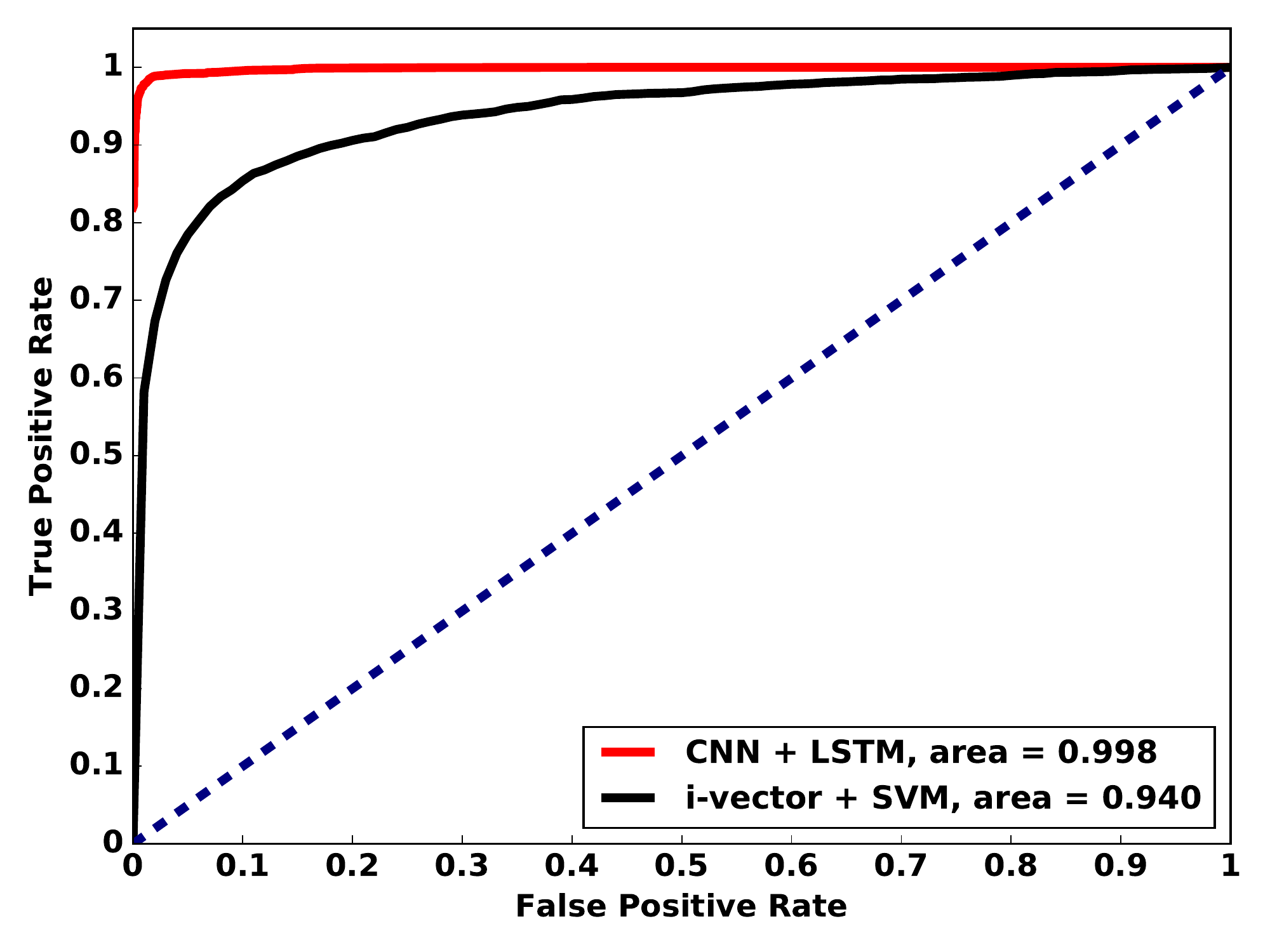}
\end{minipage}\hfill
\caption {\textbf{Top:} Speaker identification performance with change in i-vector dimensionality for four different classifier settings. Bottom: ROC curves for speaker identification with breath using CNN-LSTM and SVM classifiers.}
\label{fig:ivectors}
\vspace{-0.5cm}
\end{figure}

The speaker recognition accuracies obtained with different feature set dimensionalities in the i-vector based experiments are plotted in Fig. \ref{fig:ivectors}. The best accuracies obtained in each experiment were: \textbf{i-vector + SVM:} 72.8\%, \textbf{i-vector + NN (1 layer):} 73.5\%, \textbf{i-vector + NN (2 layers):} 71.9\% and \textbf{i-vector + LDA + SVM:} 74.1\%. We see that i-vectors derived from breath sounds alone are able to provide good speaker identification performance in this task, highlighting the basic fact that intervocalic breath sounds do carry information about the speaker, and can be successfully used for speaker identification.

\subsection{CNN-LSTM experiments}
The data set used for this set of experiments comprised 9376 instances corresponding to 44 speakers for the clean speech high-fidelity channel condition, for which the breath instances were in sufficient numbers to use for our experiments. To select network parameters and combat over-fitting, we used cross-validation. Specifically, for each speaker, we selected 70\% of the utterances as training set, 20\% as validation set, and 10\% as the test set. 
We first converted all the data-sets to constant-Q spectrograms  \cite{brown1991calculation} with 48 filters per octave, the sampling frequency $f_{\mathrm{s}}=44100$ Hz, the lowest frequency  $f_{\mathrm{min}} = 27.5$ Hz, and the highest frequency $f_{\mathrm{max}}=f_{\mathrm{s}}/2$. To compensate for pitch variations within speakers, we further augmented the data using the elastic transform  \cite{simard2003best} with $\sigma=2$ and $\alpha=15$. 

We then configured the network to have the following parameters -- Input dimension: $463\times T$; CNN filter size: $8\times 3 \times 3$; Max pooling stride: $2\times 1$; Dropout rate: 0.4; Output dimension: 44. The network was implemented using \textit{Theano}  \cite{2016arXiv160502688short}, and trained using Adadelta  \cite{zeiler2012adadelta} with decay constant 0.9 and batch size 1. The training was stopped when the errors on the validation set stopped decreasing.

Using this network, the speaker identification accuracy for breath on the test data was 91.3\%. The proposed CNN-LSTM speaker identification framework achieves \textit{remarkably accurate speaker identification} with constant-Q representations of short phoneme recordings. As a final test, we also used the individual breath recordings in a speaker {\em verification} framework where, for each speaker a {\em two-class} classifier was trained to distinguish between the speaker and the set of all other speakers. 
The bottom panel of Fig. \ref{fig:ivectors} shows the receiver operating curves (ROC) for both the CNN-LSTM based experiments and the i-vector based experiments using the 44-speaker data. In both cases, the area under the curve (AUC) is good enough to allow the potential use of breath in practical speaker identification tasks.

\section{Conclusions}
\label{sec:concl}
Experiments with both i-vectors and constant-Q spectrograms show that breath sounds can be successfully used to identify speakers. In fact, for the clean speech signals that we used in our experiments, the accuracies are surprisingly good. Note that since our primary objective in this paper is to demonstrate that the sound of the human breath can indeed be used for speaker identification, this choice of features was judiciously made to fulfill our goal of providing proof-of-concept. In future, our goal would be achieve the \textit{best} speaker recognition performance form the human breath, and the choice of features would be much more diverse and differently motivated for this. The space of classifier types can also be explored to a much greater extent.

The CNN-LSTM neural network based framework proposed in this paper uses constant-Q representations to effectively normalize the shifts in the resonance patterns of breath within the same speaker and to emphasize the distinction between speakers. The CNN-LSTM based approach automatically learns shift-invariant and temporal features, and combines feature extraction, speaker modeling and decision making into a single pipeline. This framework is also distribution-assumption free and works effectively for short recordings. 

\clearpage
\newpage
\pagebreak

\bibliographystyle{IEEEtran}

\bibliography{IEEEabrv,breathrefs}

\end{document}